\documentclass{article}
\usepackage[utf8]{inputenc}
\usepackage{textcomp}
\usepackage{amsmath}
\usepackage{amssymb}
\usepackage{graphicx}
\usepackage{braket}
\usepackage{url} 
\usepackage{comment} 
\usepackage{subfigure}
\usepackage[makeroom]{cancel}
\usepackage{xcolor}
\usepackage{mathtools}
\usepackage{authblk}

  \usepackage[
    backend=biber,
    style=ieee,
  ]{biblatex}

\usepackage[bblfile=main]{biblatex-readbbl}

\begin{document}

\title{Is Bohmian mechanics missing some motion? \\Why a recent experiment is inconclusive.}

\author[1,a,*]{Mordecai Waegell}

\affil[1]{\small{Institute for Quantum Studies, Chapman University, Orange, CA 92866, USA}}

\affil[a]{ORCID: 0000-0002-1292-6041}

\affil[*]{Corresponding Author, email: waegell@chapman.edu}

\date{\today}

\maketitle

\begin{abstract}
A recent experiment raises a supposed challenge to Bohmian mechanics, claiming to observe stationary states, which should have zero Bohm velocity, while indirectly measuring a nonzero speed based on how an evanescent wavefunction spreads from one waveguide to another coupled waveguide.  There were numerous problems with this experiment and how it was interpreted.  First, the experiment is not observing stationary states as claimed, but rather the time-averaged density of wave pulses which reflect off the potential step.  Second, the proposed method for measuring a propagation speed is shown to be invalid for true stationary states.  Third, the invalid method was misapplied to the time-averaged density, and this is shown to have created the false impression that it yields correct speed values for stationary states.  These issues notwithstanding, for a wavefunction $\psi = Re^{iS/\hbar}$, the velocity of interest, $\vec{v}_s = -\frac{\hbar}{m}\frac{\vec{\nabla}R}{R}$, is different from the Bohm velocity $\vec{v}_B=\frac{1}{m}\vec{\nabla}S$, and may be nonzero for stationary states.  This quantity has been called the \textit{symmetric} or \textit{osmotic} velocity, and while it emerges naturally as an imaginary component of the velocity in the derivation of the Madelung/Bohm description, it is usually disregarded.  So, even though we do not think this experiment makes a compelling case for it, if $\vec{v}_s$ is somehow associated with real physical motion, then this motion is indeed absent from Bohmian mechanics, as the authors contend.  We discuss a generalized Madelung fluid model where this velocity is given physical meaning, and show how it roughly agrees with the authors' concept of an evanescent De Broglie speed. 
     \end{abstract}

     \textbf{Keywords:} evanescent fields, quantum tunneling, Bohm/Madelung mechanics, quantum well, quantum optics
\pagebreak

\section{Introduction}
As discussed in the abstract, a recent article \cite{sharoglazova2025energy} raises a supposed challenge to Bohmian mechanics \cite{bohm1952suggested}, and argues for the existence of nonzero speeds in stationary eigenstates of the 2D Schr\"{o}dinger equation, where Bohmian mechanics predicts speed zero.  The experiment used an optical setup to simulate 2D Schr\"{o}dinger dynamics in the paraxial approximation.  We unpack the list of problems with this experiment and its interpretation here.

First, their experiment did not actually excite stationary states, but rather localized pulses, which are created at rest, accelerate down a ramp and then propagate along a waveguide in the main well, before reflecting off the end of a step, where the evanescent wavefunction of interest is present in a pair of coupled waveguides.  These wave pulses have nonzero Bohm velocity in the evanescent region, which undermines the core claim of the article.  This was also pointed out in \cite{drezet2025comment, daem2025trajectories}.  The stationary patterns that the authors are actually observing are the time-average of this wave pulse motion, which is clear from how the intensities are observed.  These patterns are not the stationary bound eigenstates of the well, which truly do have zero Bohm velocity, and energy eigenvalue $E_n$.  Thus, nothing we infer from the observed spatial intensity pattern has anything to do with any stationary eigenstates, so even if the speed the authors claim to measure in this way has some physical meaning (and it does seem to be consistent with the energy expectation value $\langle E\rangle$ of the pulse), it is unrelated to any eigenstate, even if its energy eigenvalue $E_n = \langle E\rangle$.

Second, the method the authors are using to supposedly measure the propagation speed in stationary eigenstates is shown to be invalid using numerical solutions for the 2D bound eigenstates of the well.  For each eigenstate, we computed the evanescent De Broglie speed $\sqrt{2(V-E)/m}$, as well as the velocity $-\frac{\hbar}{m}\frac{\vec{\nabla}R}{R}$, and evaluated them just inside the surface of the step, where they were found to be in good agreement.  We then applied the authors' method for determining a propagation speed in the evanescent region just inside the surface of the step based on the relative rate of spatial population spreading between the main and and auxiliary waveguides, and we found that this value was in poor agreement with the first two, especially for lower energies.  This seems to show relatively conclusively that the proposed method for measuring the propagation speed does not work - at least when applied to the truly stationary eigenstates for which it was intended.

Third, two of the previous errors compounded to create another.  Thinking their time-averaged intensity patterns were for stationary states of energy eigenvalues $E_n = \langle E \rangle$, the authors applied their method, and found propagation speeds consistent with the $\sqrt{2(V-\langle E \rangle)/m}$.  From this, they mistakenly concluded that their method is a valid way of measuring $\sqrt{2(V-\langle E \rangle)/m}$ and $-\frac{\hbar}{m}\frac{\vec{\nabla}R}{R}$ for stationary eigenstates, and that indeed they had successfully measured these speeds in these stationary states.

While the results and interpretation of this experiment are not compelling, the velocity $-\frac{\hbar}{m}\frac{\vec{\nabla}R}{R}$ they are purporting to measure, which has been called the \textit{symmetric} \cite{waegell2024toward,waegell2024madelung} or \textit{osmotic} \cite{nelson1966derivation, bohm1989non, nelson2012review} velocity, is not generally zero in stationary states, and this may yet be of interest.  If this motion is physically real in some way, it is indeed true that it is absent from Bohmian mechanics.  However, as we will discuss, this quantity does appear naturally as an imaginary component of the velocity in deriving the Madelung \cite{madelung1927quantum}/Bohm equations \cite{bohm1952suggested}.  This has also led some authors to develop a treatment called Bohmian Mechanics with Complex Action (BOMCA) involving complex velocities and in complex position space \cite{poirier2012action}.  This same quantity arises from random fluctuations is Nelson's derivation \cite{nelson1966derivation} - hence \textit{osmotic}), and in the \cite{waegell2024toward,waegell2024madelung} it relates to local velocity distributions which sum to zero - hence \textit{symmetric}.

That said, we should stress that Bohmian mechanics, without any ontological interpretation of the symmetric velocity, already makes the same empirical predictions as standard quantum mechanics, including the results of this experiment.  Nevertheless, there are some other tantalizing reasons, discussed later, to suppose it might have an ontological status in other empirically consistent interpretations - particularly the local many worlds model \cite{waegell2023local}, where the same equations describe many fluids in 3-space, building on Madelung \cite{madelung1927quantum,waegell2024toward,waegell2024madelung}.

The remainder of this article is organized as follows.  In Sec. \ref{Velocity}, the origin of the Bohm and symmetric velocity with the mathematical formalism, and discuss reasons for thinking there is physical motions associated with the symmetric velocity.  In Sec. \ref{Sim} we discuss the details of our numerical simulation of the experiment.  In Sec. \ref{Tuning} we discuss how the experimental potential is approximated and tuned to match the parameters in the experiment.  In Sec. \ref{Stat} we examined the  bound stationary eigenstates of this potential, and evaluate the De Broglie speed, the symmetric speed, and authors' method of inferring speed based on the population transfer between the waveguides.  In Sec. \ref{Pulse} we simulated wave pulses, construct time-average densities, and used the authors' method to infer speed values.  We end with some discussion in Sec. \ref{Conc}.

\section{The Bohm velocity and the symmetric velocity} \label{Velocity}

The authors seem to recognize that the velocity they are studying, $\vec{v}_s = -\frac{\hbar}{m}\frac{\vec{\nabla} R}{R}$, is not the same physical quantity as the Bohm velocity, $\vec{v}_B = \frac{1}{m}\vec{\nabla}S$, where $R$ and $S$ come from the polar (or eikonal) representation of the wavefunction $\psi = Re^{iS/\hbar}$.  The symmetric velocity $\vec{v}_s$ comes from the imaginary part of the momentum density $\psi^*\hat{p}\psi$ in the derivation of Bohm or Madelung mechanics, and is closely related to the \textit{quantum potential}, while the standard Bohm velocity $\vec{v}_B$ comes from the real part.  The symmetric velocity $\vec{v}_s$ has been developed within a generalized (local) Madelung fluid model \cite{waegell2024toward,waegell2024madelung,waegell2023local} (which uses the same mathematical formalism as Bohmian mechanics), and is known to be nonzero in stationary eigenstates, where the associated kinetic energy is roughly consistent with the energy eigenvalue of the state.  In this model, the symmetric velocity is understood as a directionless contribution similar to what the authors suggest, but this model emerges more naturally from the Madelung equations than the bidirectional model presented in their followup note reaffirming the challenge to Bohm \cite{klaers2025reaffirming}.

What are the reasons to think the symmetric velocity might be associated with real physical motion?  On the one hand, the fact that eigenstates are stationary in Bohmian mechanics seems to have the conceptual advantage that it may seem to explain why bound charges do not radiate all of their energy away as classical electromagnetic radiation - however degenerate superpositions with nonzero angular momentum can have nonzero Bohm velocity while still failing to radiate, so this explanation does not work.  On the other hand, it is counterintuitive that bound states whose classical analog would definitely be in motion should be at rest in the quantum case, and there is experimental evidence that muons in bound states whose symmetric velocity is relativistic display a corresponding time-dilation of the their half-lives \cite{silverman1982relativistic}, which seems to suggest motion.  In Bohmian mechanics, the particles are at rest, and all of this energy is stored in the \textit{quantum potential}.

So, we agree with the authors that there is compelling reason to think that there is motion associated with $\vec{v}_s$, but we think one must be very careful in interpreting this motion.  In particular, we find the present argument based on the spatial spreading rates of the evanescent field, and the associated model described in \cite{klaers2025reaffirming}, to be dubious, and not supported by the experimental evidence. 

Whatever motion $\vec{v}_s$ represents, it does not contribute in any way to the dynamics of the total Madelung fluid density $\rho \equiv R^2 \equiv |\psi|^2$, which is fully characterized by $\vec{v}_B$.  Thus, if $\vec{v}_s$ does describe flow, it must be such that the net flow into and out of any region due to $|\vec{v}_s|$ is zero.  The meaning of the direction of $\vec{v}_s$ is not entirely clear, but one proposition is that it is the axial direction for infinitesimal orbits at speed $|\vec{v}_s|$ \cite{waegell2024madelung}.

When we simulate a Gaussian wave pulse with energy above the barrier height, the population transfer between the waveguides is fully explained by $\vec{v}_B$.  And when we simulate a fully bound pseudo-Gaussian (defined later) pulse reflecting off the barrier, the spreading into the auxiliary waveguide as the pulse arrives at the barrier, and the squeezing back together into the main waveguide as the pulse departs are also fully explained by $\vec{v}_B$ (this was shown in detail in \cite{daem2025trajectories}).  The comparison with dwell time is meaningful in this case, but it does not appear to make much sense in the stationary case.

So while there is spatial spreading in the evanescent solutions for stationary eigenstates, and the symmetric velocity $\vec{v}_s$ is nonzero, we think it is unlikely to be explained by trajectories that actually spread as they travel down the waveguide, and then reverse and squeeze back together on their way out.

\section{Numerical Simulation} \label{Sim}

We have numerically solved the 2D Schr\"{o}dinger equation for a potential which closely approximates the one used in the experiment, and using their effective particle mass of $m = 6.95 \times 10^{-36}$ kg.

This results in 103 bound eigenstates $\psi_n$, with energy eigenvalues $E_n$ ($n=1,\ldots,103$), which are nonseparable (i.e., $\psi_n(x,y) \neq f_n(x) g_n(y)$).  However, because the cavity is long and narrow, they still approximate the separable eigenstates of the a harmonic waveguide with walls at the ends, and 72 of the eigenstates are in the harmonic oscillator ground state with respect to the transverse ($y$) direction, $n_y \approx 0$, while 31 are in the first excited state with $n_y \approx 1$.  For the 72 with $n_y \approx 0$, the bound states have $n_x \in \{1,\ldots 72\}$, and for the 31 with $n_y \approx 1$,  they have $n_x \in \{1,\ldots 31\}$.  The energy eigenvalues also break down approximately as $E_n \approx E_{n_x} + E_{n_y}$ (there is no particular relation between the true quantum number $n$, and the approximate pseudo-quantum-numbers $n_x$, and $n_y$).

Furthermore, the energy cutoff for bound states is the $y$-direction ground state energy on the raised step, $E^\textrm{step}_{y_0}$, meaning all 103 eigenstates with lower energy than this are bound to the well, with exponentially decaying tails in the step, and all energy eigenstates with higher energy than this are free.

The numerical simulation was done using MATLAB's `eigs' function for a Hamiltonian set up on a mesh with $-900\leq x\leq700$ $\mu$m, with a spacing $h_x = 0.5$ $\mu$m, and $-40\leq y \leq 40$ $\mu$m, with spacing $h_y = 0.1$ $\mu$m.  

\subsection{The Potential Step and Coupling Constant} \label{Tuning}

In order to match the experiment, we tuned our simulated potential in two ways.  The most important parameter to match seems to be the coupling constant between the main and auxiliary waveguides, which was determined in their experiment to be $J_0 \approx 2\pi(6.34)$ GHz, which is related to the rate of population transfer between the coupled waveguides on the raised step.  To an excellent approximation, this frequency is given by the two lowest $y$-eigenvalues in the coupled waveguides as $J_0 = \frac{1}{2\hbar}(E^\textrm{step}_{y_1} - E^\textrm{step}_{y_0})$, since the population transfer along $y$ is due to the standard time evolution of an equal superposition of the ground state and first excited state (see also \cite{daem2025trajectories}).  The basic shape of the $y$ potential was chosen to approximate the surface height of the nanostructured mirror in the experiment, and then an overall scaling factor $V_s$ was tuned so that the resulting $J_0$ matches the experimental values.

The authors use a pair of coupled 1D Schr\"{o}dinger equations along $x$ as an \textit{ansatz} for the true 2D solution, so in order to get comparable results, we need to properly isolate the motion along $x$.  Thankfully, the fact that these modes are approximately separable makes this straightforward.  The effective barrier height for $x$ motion is given by the difference between the $y$ ground state energies in the well and on the step, $V_0 = E^\textrm{step}_{y_0}-E^\textrm{well}_{y_0}$, which follows from assuming that the total particle energy in the well, $E^\textrm{well}_x + E^\textrm{well}_{y_0}$ (assuming the ground state energy in $y$), must exceed $E^\textrm{step}_{y_0}$ in order for the particle to move onto the step.  

With $J_0$ already fixed, we can next tune the well-floor-to-step-floor height $h_0$ to produce an effective potential step in $x$ that matches the experimentally determined value $V_0 \approx 0.538$ meV $= 0.862 \times 10^{-22}$ J.  The relevant energy eigenvalues are, $E^\textrm{well}_{y_0} = 0.205$ meV $E^\textrm{step}_{y_0} = 0.743$ meV, and $E^\textrm{step}_{y_1} = 0.796$ meV, and these are found by numerically solving the 1D Schr\"{o}dinger equation using the $y$-potentials in the well, and on the step, respectively.

\begin{figure}
    \centering
    \includegraphics[width=\linewidth]{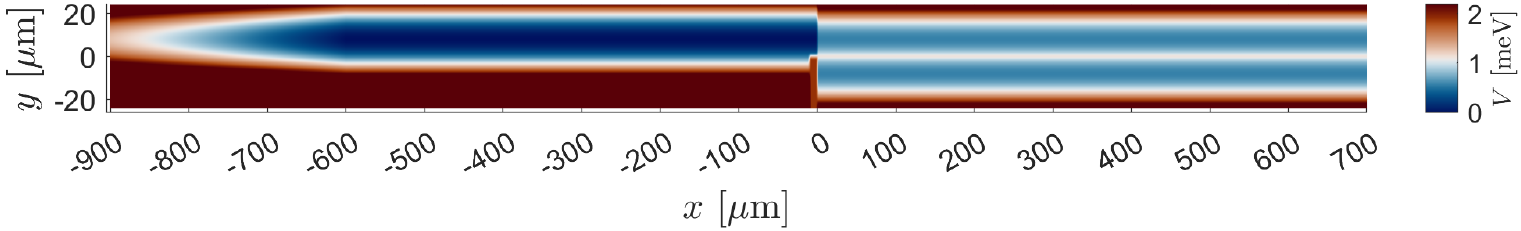}
    \caption{The potential used for our simulation, which is designed to closely mimic the one used in the experiment, shown roughly to scale.  We cut off the highest values of the potential here so that the color map provide a more visible contrast.  We used a similar color map to the one used in Fig. 1b of the article to aid comparison, but it is not quite a perfect match.}
    \label{V}
\end{figure}

\begin{figure}
    \centering
    \includegraphics[width=\linewidth]{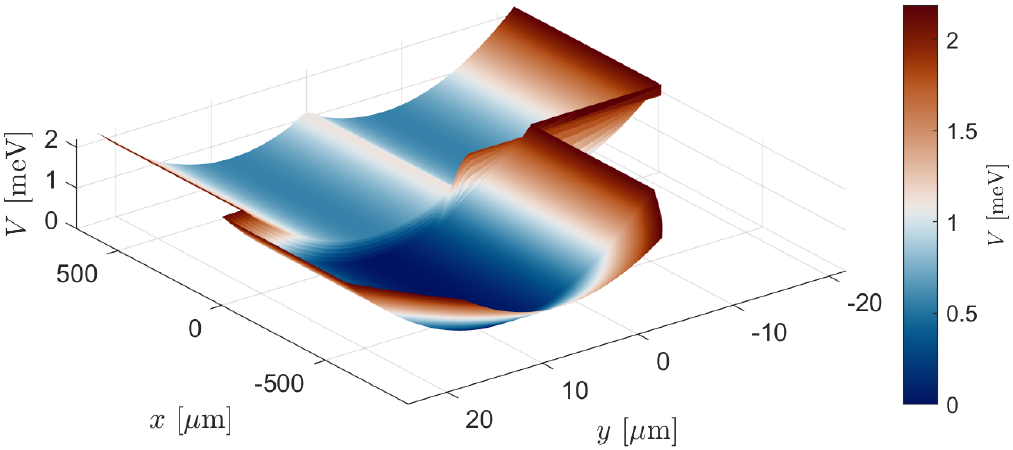}
    \caption{3D image of the smoothed potential energy surface used in the simulation.}
    \label{V2}
\end{figure}

Our potential is then constructed piecewise as follows,
\[V(x,y) = V_s\begin{cases}
     \frac{4}{75 \textrm{ }\mu\textrm{m}^2}(y - 8\textrm{ }\mu\textrm{m}  )^2 -   \frac{2}{75 \textrm{ }\mu\textrm{m}}x-16& -900 \textrm{ }\mu\textrm{m}\leq x< -600 \textrm{ }\mu\textrm{m} \\

    \frac{4}{75 \textrm{ }\mu\textrm{m}^2}(y - 8\textrm{ }\mu\textrm{m}  )^2 & -600 \textrm{ }\mu\textrm{m} \leq x< -10 \textrm{ }\mu\textrm{m} \\

    \frac{4}{75 \textrm{ }\mu\textrm{m}^2}(y - 8\textrm{ }\mu\textrm{m}  )^2 &-10 \textrm{ }\mu\textrm{m} \leq x < 0  \textrm{, }y>0 \\
    
    12 &-10 \textrm{ }\mu\textrm{m} \leq x< 0  \textrm{, }y\leq0 \\

    \frac{4}{75 \textrm{ }\mu\textrm{m}^2}(y - 8\textrm{ }\mu\textrm{m}  )^2 + h_0 & 0 \leq x\leq 700 \textrm{ }\mu\textrm{m} \textrm{, }y> 0 \\

        \frac{4}{75 \textrm{ }\mu\textrm{m}^2}(y + 8\textrm{ }\mu\textrm{m}  )^2 + h_0 & 0 \leq x\leq 700 \textrm{ }\mu\textrm{m} \textrm{, }y\leq 0 ,\\
   \end{cases}
\]
with $V_s = 0.1581$ meV, and $h_0 = 3.587$.  The first row is the ramp in the $x$ direction, and a harmonic potential in the $y$ direction.  The second row is the main well, which is a free waveguide in $x$ and harmonic in $y$. The third row is harmonic in $y>0$.  The fourth row is a flat barrier that limits direct spread from the main well into the auxiliary waveguide on the raised step.  The fifth row is the main waveguide in $x$, with a harmonic potential in $y>0$, and the sixth row is the auxiliary waveguide in $x$ with a different harmonic potential in $y\leq 0$.  To make the simulation more physical, all of the piecewise transitions in $V(x,y)$ are smoothed using a cosine-weighted average that extends 2 $\mu$m to either side of the boundary.  This smoothed 2D potential surface is depicted in Figs. \ref{V} and \ref{V2}.  The major difference between this potential and the one used in the experiment is that we used harmonic potentials for the waveguides instead of finite walls, but they are closely matched in the energy regime of interest, and should have nearly the same bound eigenstates.

\subsection{Simulated Speeds for Stationary Energy Eigenstates}  \label{Stat}

To compare with the 1D ansatz, we need to isolate the $x$ part of the energy, which we do by assuming that the states have the ground state $y$ energy both in the well and on the step.  The total 2D energies are $E = E_x + E^\textrm{well}_{y_0}$, and the true barrier potential in 2D is $E^\textrm{step}_{y_0} = V_0 + E^\textrm{well}_{y_0}$, so we get our effective 1D picture by subtracting $E^\textrm{well}_{y_0}$ from both, and our the 1D analog particle has energy $E_x$ and encounters a potential barrier of height $V_0$.

\begin{figure}
    \centering
    \includegraphics[width=0.9\linewidth]{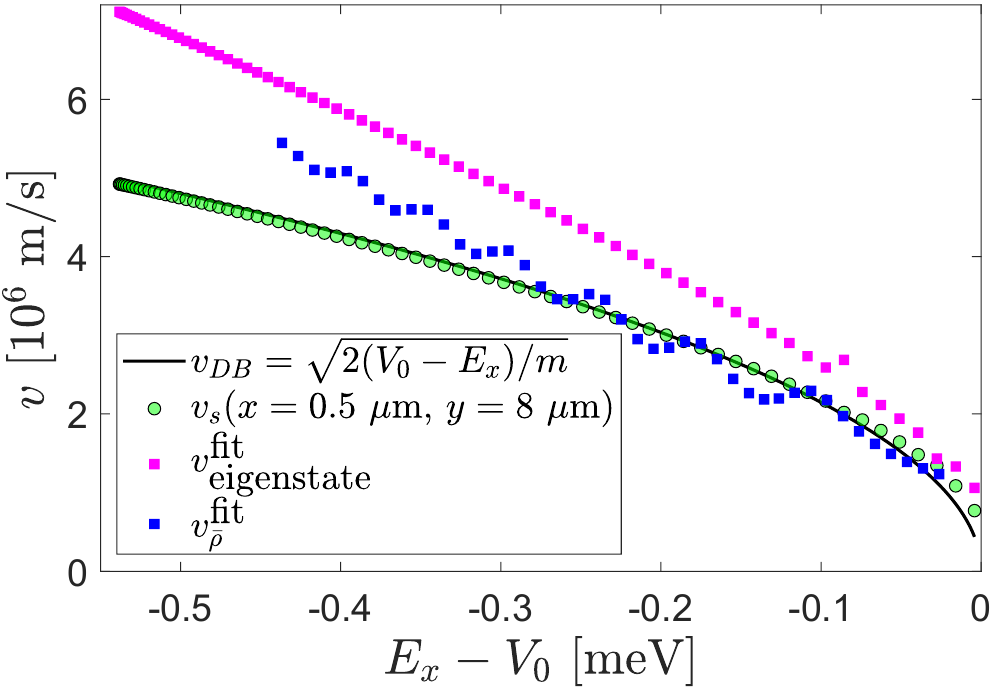}
    \caption{Evanescent speeds, just inside the step, using an effective 1D treatment for comparison with \cite{sharoglazova2025energy}.  $v_{DB}$ is the so-called De Broglie speed for a particle with $x$-energy $E_x$.  For the 72 eigenstates with $n_y=0$, the of symmetric speed $\vec{v}_s = -\frac{\hbar}{m}\frac{\vec{\nabla} R}{R}$ is computed and evaluated at the center of the main waveguide ($y = 8$ $\mu$m).  For the same 72 eigenstates, the relative centerline population of the auxiliary waveguide, $\rho_a(x) \equiv |\psi_a(x)|^2/[|\psi_m(x)|^2 + |\psi_a(x)|^2]$ $= |\psi(x,-8$ $\mu$m$)|^2/[|\psi(x,8$ $\mu$m$) + |\psi(x,-8$ $\mu$m$)|^2]$ for $0<x<10.5$ $\mu$m, was fit to the function $Cx^2$, from which the inferred propagation speed is $v_{\textrm{eigenstate}}^{\textrm{fit}} = J_0/\sqrt{C}$ (the $n_x = 65$ and $n_x=70$ eigenstates are poorly approximated as separable, and do not fit the same pattern as the other $n_x$ values).  Wave pulse reflections were simulated at 42 different average energies, separated by roughly $0.01$ meV, and time-averaged densities $\bar{\rho}(x,y) = \frac{1}{T}\int_T |\psi(x,y,t)|^2 dt $ were computed.  For these 42 cases, the relative centerline population of the auxiliary waveguide, $\rho_{\bar{a}}(x) \equiv \bar{\rho}(x,-8$
    $\mu$m$)/[\bar{\rho}(x,8$ $\mu$m$) + \bar{\rho}(x,-8$ $\mu$m$)]$ was fit to function $Cx^2$, from which the inferred propagation speed is $v_{\bar{\rho}}^{\textrm{fit}} = J_0/\sqrt{C}$.
    }
    \label{Speeds}
\end{figure}

Our numerical simulation shows good agreement between the so-called evanescent De Broglie speed, $v_{DB} = \sqrt{2(V_0 - E_x)/m}$ and the value $\vec{v}_s$ just inside the edge of step in the center of the main waveguide, but these give poor agreement with velocity determined using the authors' method involving $J_0$ and the parabolic fits to the spreading rates.  There is a clear dependence between the spreading rate and the $x$ energy eigenvalue $n_x$, with distinct series for $n_y=0$ and $n_y=1$, but it does not match the form predicted be the authors.  In Fig. \ref{Speeds}, we focus on the $n_y=0$ series, and plot the three different speeds values obtained at energy $E_{n_x}$ for $n_x = 1,\ldots, 72$, and the discrepancy is largest for small values of $n_x$.    

There is a subtle inconsistency between our 2D numerical solutions and the 1D \textit{ansatz} used by the authors \cite{klaers2023particle, sharoglazova2025energy}, which presents a slight problem for the comparison.  The 2D numerical solutions don't seem to have an intermediate `long range non-oscillatory' region (see Fig. 1(c) of \cite{sharoglazova2025energy}), so we associate this with bound states in the well that are just below the threshold to be free, since these cases still appear to exponentially decay at a slow rate.  As such, we tuned $V_0$ so that states with $E_x < V_0$ are bound in the well, and thus states with  $\Delta \equiv E_x - V_0 + \hbar J_0 < \hbar J_0$ are bound.  However, in the \textit{ansatz} solution the De Broglie speed is given by $\tilde{v}_{DB} = \sqrt{2|\Delta|/m}$ (see Fig.3(a) of \cite{sharoglazova2025energy}), which is notably different from the ${v}_{DB} = \sqrt{2(V_0 - E_x)/m}$ we get from our tuning, and has its zero in a different place.  This relation seems to indicate that the threshold between bound and free states is at $\Delta = 0$, so that states with $E_x < V_0 - \hbar J_0$ are bound in the well.  Thus, there seems to be a discrepancy regarding the effective barrier height within the \textit{ansatz} model itself.  This is likely related to the fact that its solutions seem to show a third intermediate region of behavior between classically allowed (oscillatory) and classically forbidden (exponential), while our 2D numerical solutions do not seem to show any such intermediate region.  Note that if only the states the authors call `exponentially decaying' ($\Delta<-\hbar J_0$) are bound, then the threshold is $E_x < V_0 - 2\hbar J_0$.

\subsection{Simulated Speeds for Time-Averaged Pulse Densities} \label{Pulse}

The experiment was performed by inserting wave pulses at rest at different locations on the ramp in order to get different kinetic energies in the main well, once the pulse has accelerated down the ramp.  Despite this clearly dynamical approach, the authors claim that the density data they are collecting are revealing the shapes of stationary energy eigenstates.  This is incorrect, which can be seen in several ways.  The simplest is that the energies used do not seem to match the energy eigenvalues of the actual well.  But the main issue is that the imaging technique in use clearly reveals the time-averaged density of the wave pulses as they propagate and reflect off the step, and there is no reason to expect these to be equivalent to eigenstate densities.

One can understand how the authors might have made this mistake: the time average-densities look very much like standing wave patterns because even as the wave propagates and reflects, the positions of the interference fringes remain nearly stationary - and this is true regardless of the average energy of the pulse.  The authors used these patterns to determine the $x$ energy of the prepared pulses, which yields correct values for the expectation value $\langle E_x\rangle$.

We simulated the experiment by creating pseudo-Gaussian pulses at rest at different locations on the ramp.  We begin with a separable Gaussian pulse $\phi_0(x,y) = f(x)g(y)$ with $g(y)$ closely matching the harmonic oscillator ground state in the $y$-direction with a standard deviation $\sigma_y \approx 4.92$ $\mu$m, and for $f(x)$, we use a Gaussian twice as wide so that the decomposition into the 103 bound states is closer to converging, but the positions on the ramp is still somewhat localized.  We then approximated the Gaussian pulse using the renormalized decomposition into those 103 bound states, which guarantees that there is no transmission over the step.
Explicitly, we use numerical integration to find the coefficients,
\begin{equation}
    c_n = \iint \phi_0(x,y) \psi_n(x,y) dx dy,
\end{equation}
for the bound eigenstates ($\psi_n(x,y)$ with energy eigenvalues $E_n$) for $n=1,\ldots,103$, and then renormalize this set to obtain, $c_n' = c_n/\sqrt{\sum_{n=1}^{103}|c_n|^2}$, from which we construct the state pseudo-Gaussian $\phi_0'(x,y) = \sum_{n=1}^{103}c_n'\psi_n(x,y)$.

In this way, we constructed 42 test pulses $\phi_0'$ with different average energies $\langle E\rangle = \sum_{n=1}^{103}|c_n'|^2 E_n$, separated by roughly $0.01$ meV.  This results in a pulse $\phi_0'$ that is mostly localized near $x_0$ on the ramp, but for energies $E$ closer to the barrier height $V_0'$, the fidelity $|\langle \phi_0|\phi_0'\rangle|^2$ with which the fully bound $\phi_0'$ approximates the intended Gaussian pulse $\phi_0$ gets worse, and there are also some wavelike components whose amplitude decays away from the pulse.  This suggests that the pulses being used in the experiment may have been exciting modes above the barrier energy, although the experimental data does not seem to show this.

For each of the test pulses, we simulated the time-evolution of the pulse as it reflects off the barrier, and constructed the time-averaged density $\bar{\rho}$ over that process.  Next, we evaluate the rate of population spreading into the auxiliary waveguide in each of these time-average densities, and following the authors, perform parabolic fits along with $J_0$ to obtain a speed value, as shown in Fig. \ref{Speeds}.

Remarkably, at least for a range of energies close to the barrier height, the speeds obtained in this way seem to roughly match the De Broglie speed $\sqrt{2(V_0-\langle E_x \rangle)/m}$.  This explains what was actually observed in the experiment, and how it tricked the authors into thinking they were successfully measuring $\sqrt{2(V_0-E_{n_x} \rangle)/m}$ and $\vec{v}_s$ for a stationary eigenstate with $E_{n_x} = \langle E_x \rangle$.  It seems plausible that the apparent agreement is actually due to the time-averaged effect of the Bohm velocity as the pulse actually moves into and out of the barrier during the reflection.

Note that despite the apparent matching at various energies, we also cannot meaningfully compare the 42 speeds obtained in these fits to the $\vec{v}_s = -\frac{\hbar}{m}\frac{\vec{\nabla} R}{R}$ for 72 eigenstates, since for the time-average densities do not match the eigenstate densities, even if $E_{n_x} = \langle E_x \rangle$.

\section{Conclusions}\label{Conc}

Whether or not the symmetric velocity has an interpretation as physical motion remains an interesting issue.  If it does, then the generalized Madelung model we have been discussing may explain the motion that is missing from Bohmian mechanics.

That said, we think we have shown that this experiment does not measure what the authors intended, and does not provide compelling evidence of this missing physical motion, so there does not seem to be a significant challenge to Bohmian mechanics at present.  

This case should also serve as a cautionary tale, where a confluence of error and coincidence created the convincing illusion of a direct physical relation between evanescent spatial spreading rates and symmetric velocity in stationary eigenstates.

Even though it is not the one the authors expected, and may not indicate the trajectories they were envisioning, Fig. \ref{Speeds} shows that there is still some approximate relation between spatial spreading rates and the energy eigenvalues $E_{n_x}$ and $E_{n_y}$ of stationary states.  Understanding this relation, and other related properties of evanescent fields, may have interesting applications.\newline

\noindent\textbf{Acknowledgments:} Thanks to Justin Dressel for helpful discussions.\newline

\noindent\textbf{Funding:} This project/publication was made possible through the support of Grant 63209 from the John Templeton Foundation. The opinions expressed in this publication are those of the authors and do not necessarily reflect the views of the John Templeton Foundation.\newline

\noindent\textbf{Conflict of Interest Statement:} No potential competing interest was reported by the author.

\printbibliography

\end{document}